\documentclass[aps, twocolumn, superscriptaddress,showpacs, pre]{revtex4-1}
\usepackage{graphicx,amsfonts,amsmath,color,amsbsy,amssymb, subfigure}
\usepackage{epstopdf}

\begin{document}
\title{ Induced stresses in quasi-spherical elastic vesicles: local and global Laplace-Young law.}
\author{G. Torres-Vargas}\email{torres@correo.cua.uam.mx} 
\affiliation{Posgrado en Ciencias Naturales e Ingenier\'ia.
Universidad Aut\'onoma Metropolitana Cuajimalpa.\\
Vasco de Quiroga 4871, 05348 Cd. de  M\'exico, MEXICO}

\author{J.A. Santiago}\email{jsantiago@correo.cua.uam.mx}
\author{G.  Chac\'on-Acosta}\email{gchacon@correo.cua.uam.mx}
\affiliation{Departamento de Matem\'aticas Aplicadas y Sistemas\\ 
Universidad Aut\'onoma Metropolitana Cuajimalpa\\
Vasco de Quiroga 4871, 05348 Cd. de  M\'exico, MEXICO}

\vspace{10pt}


\begin{abstract}
On elastic spherical  membranes,  there is no  stress induced by the bending energy and the corresponding Laplace-Young law  does not involve the  elastic bending stiffness. 
However, when considering an axially symmetrical perturbation that pinches the sphere, it induces nontrivial stresses on the entire membrane.
In this paper we introduce a theoretical framework to examine the stress induced by perturbations of geometry around the sphere. We find the local balance force equations along the normal direction to the vesicle, and along the unit binormal, tangent to the membrane;  likewise, the global balance force equation on closed loops is also examined. 
We analyze the  distribution of stresses on the membrane as the budding transition occurs. 
For closed membranes we obtain the modified Young-Laplace law that appears as a consequence 
of this perturbation.

\smallskip

\end{abstract}


\maketitle

\section{Introduction}

Many of the cellular processes such as morphogenesis or cell division, migration and other physical and biochemical events are determined by changes in the shape of the cell membrane, which in turn are regulated by mechanical stress and surface tension\cite{diz,clark,pontes}. The understanding of these transformations is also useful for the diagnosis of diseases since it has been seen that the membrane conformation changes during an infection\cite{park}. 
In addition to biological membranes the study of the forces on synthetic vesicles has applications in industrial encapsulation, drug delivery\cite{kunitake,elani,joseph}, colloids science, in several areas of physics, further to the computational tools and algorithms that have been developed for the study of this deformations \cite{yuan,feng}. Hence the importance of studying them and seeing vesicle curvature as a main actor in the resulting conformations\cite{gallop}.
So then, the main physical forces involved in the deformation processes of the vesicle are tension, pressure and stiffness. However, due to the difference in measurements in various experiments, the definition of the effective surface tension of the vesicle has been recently discussed\cite{gueguen},  finding modifications to the well-known Laplace law for the pressure difference through the membrane\cite{fournier,galatola}.

\begin{figure}[th]
\centering 
\subfigure[$\epsilon =0.2$]{
	\includegraphics[width=0.85in]{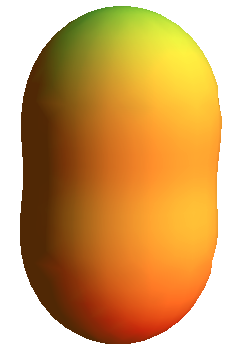}}
\hspace{- 0.50mm}
\subfigure[$\epsilon =0.6$]{
	\includegraphics[width=0.85in]{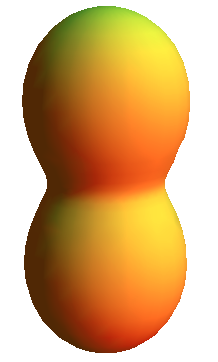}}
\subfigure[$\epsilon =0.8$]{
	\includegraphics[width=0.85in]{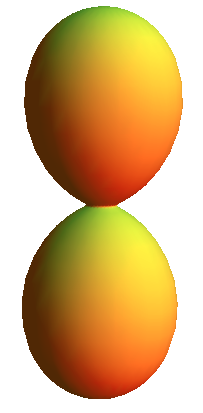}}
	\vspace{-2mm}
\caption{Sequence of the budding-like transition $f(\theta)= \epsilon P_2 (\theta) $, starting form a unit sphere and varying the parameter $\epsilon=0.2, 0.6, 0.8$. The waist appears 
at $\theta=\pi/2$.}
\vspace{-2mm}
\label{APP}
\end{figure}

It is well known that on a spherical vesicle with no spontaneous curvature there is no stress due to the bending energy. This implies a relationship between the pressure difference $P = P_{in} - P_{out}$, the surface tension $\sigma$ and the radius of the membrane $R$, given by the Young-Laplace law, $2\sigma / R = P$, which does not involve the bending stiffness $\kappa$ of the membrane. This means that in equilibrium the force (per unit length) is completely tangential to the spherical membrane and has magnitude $\sigma$, that is balanced with a force of magnitude $PR / 2$ in the opposite direction. Thus, the  pressure difference $P$ is constant along the membrane. 
Due to their composition and properties of the environment to which they are exposed, biological membranes prefer to curve in a certain specific way which is described through its spontaneous curvature. This property is essential to understand the morphology of organelles and other cellular processes\cite{rozycky}.
If the spontaneous curvature $K_0$ is nonzero the force remains tangential but now involves a coupling with the bending 
stiffness through $-(\sigma + \kappa K_0^2/ 2) +   \kappa K_0/R$, and the
Young-Laplace law is then given by \cite{seifert}
\begin{equation}
\frac{PR}{2}= \Sigma  - \frac{\kappa K_0  }{R}, \label{LPU}
\end{equation} 
where $\Sigma=\sigma + \kappa K_0^2/2$. 
The spherical vesicle is precisely the configuration of lowest energy  
of the  Canham-Helfrich functional\cite{canham,helfrich}
\begin{equation}
{\cal H}= \sigma \int dA + \frac{\kappa}{2} \int dA (K-K_0)^2 - P\int dV, \label{CAHEL}
\end{equation}
where $dA$ is the area element, $K$ the mean curvature and   
$V$ the volume enclosed by the vesicle.  From Eq. \eqref{CAHEL} we can interpret both, the pressure jump $P$ 
and the surface tension $\sigma$ as Lagrange multipliers that fix volume and surface area, respectively. Thus, according to Eq. \eqref{LPU} the pressure difference is also constant along the membrane if the spontaneous curvature does. 
Nevertheless, any small deformation of the spherical shape induces a non-trivial stress on the membrane surface, even 
along the orthogonal direction to the membrane. Such deformations can be expressed in terms of the spherical functions $Y_{lm} (\theta, \phi) $,
with $ \theta $ and $ \phi $ the polar and azimuthal angles respectively, or in terms of
Legendre polynomials $ P_l (\theta) $ in the simplest case of axial symmetry\cite{oyang}.
An important deformation of this kind that frequently appears in biological systems is the so-called budding transition\cite{wiese,julicher,julicher1,miao}, which consists in the formation of buds from the main membrane that can be modeled by two spheres connected by narrow necks. In these systems necks were formed, for instance, when external adhesion\cite{agudo} exists as it occurs in cellular processes such as phagocytosis and endocytosis\cite{dimitrieff}.
Experimentally, the transition can be induced in lipid bilayers by changing the area-volume relation or through temperature variations\cite{kas}.
This shape also appears when translocating a fluid vesicle through a tube or a pore of smaller radius, or in micropipettes aspiration experiments which are applicable in microfluidics and in drug release\cite{galatola,golushko1,shojaei,khunpetch}.
Furthermore, a similar shape also appears when studying two superposed drops with different phases\cite{kusu,li}.
These shapes can be parameterized precisely in terms of axially symmetric functions $P_2(\theta) $ as depicted in Fig. \ref{APP}: 
At the beginning of the process, the spherical vesicle is deformed at the middle, gradually a waist appears and the vesicle takes the form of a peanut. Deformation continues until it is segmented and a second spherical membrane appears.

\begin{figure}[th]
\centering 
\subfigure[$\epsilon =-0.8$]{
	\includegraphics[width=1.1in]{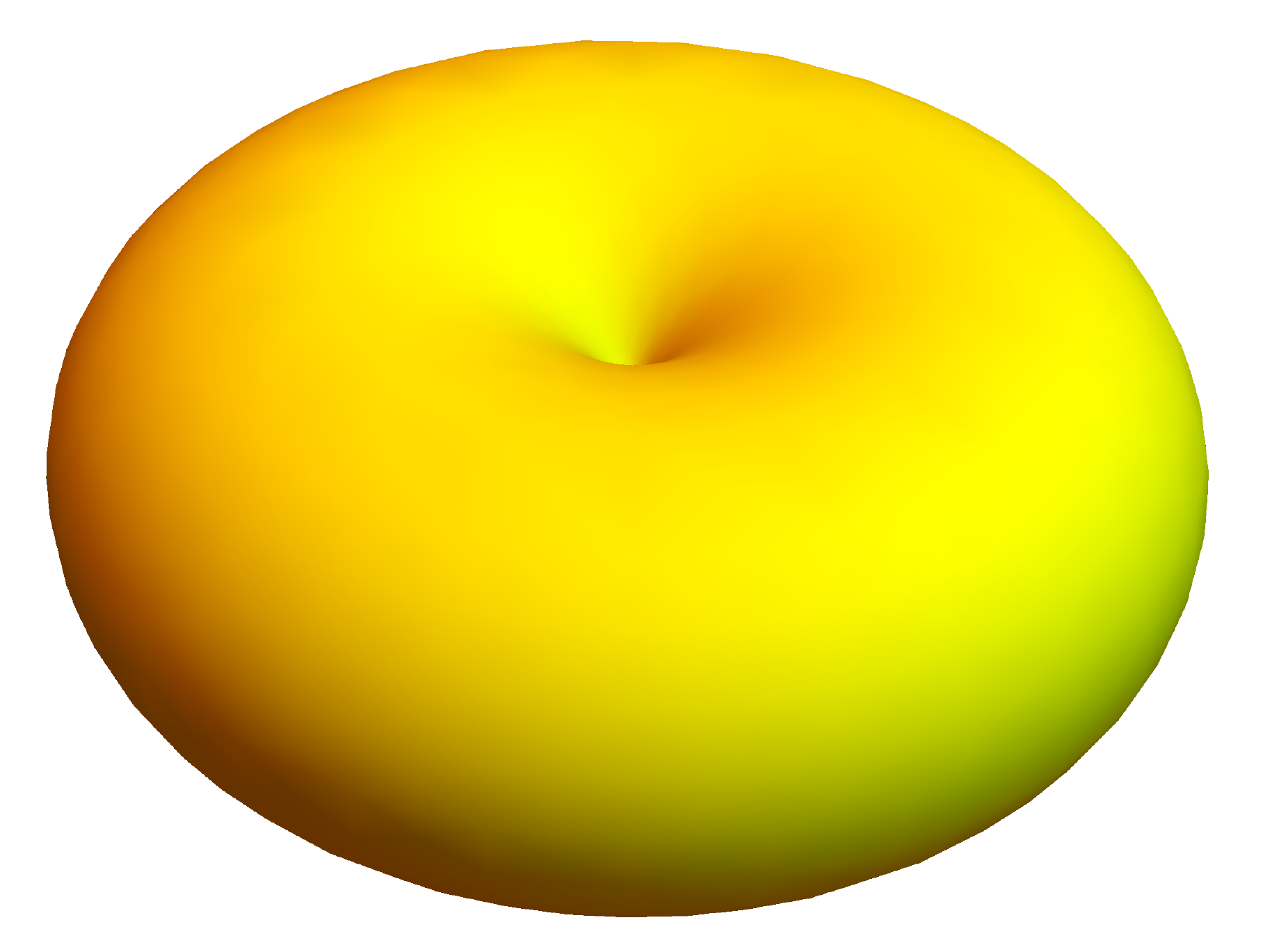}}
\subfigure[$\epsilon =-0.4$]{
	\includegraphics[width=1.1in]{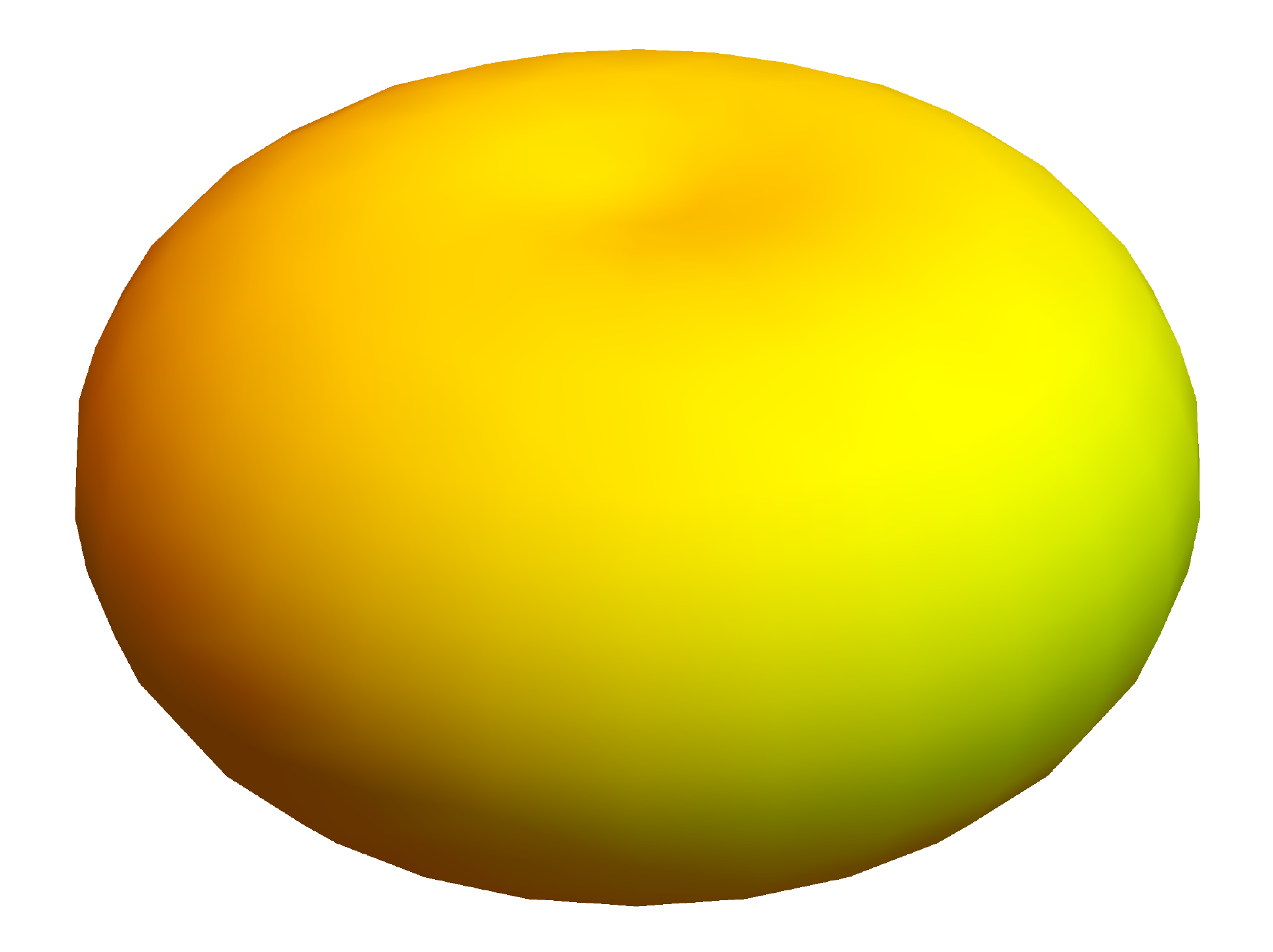}}
\subfigure[$\epsilon =0$]{
	\includegraphics[width=1.1in]{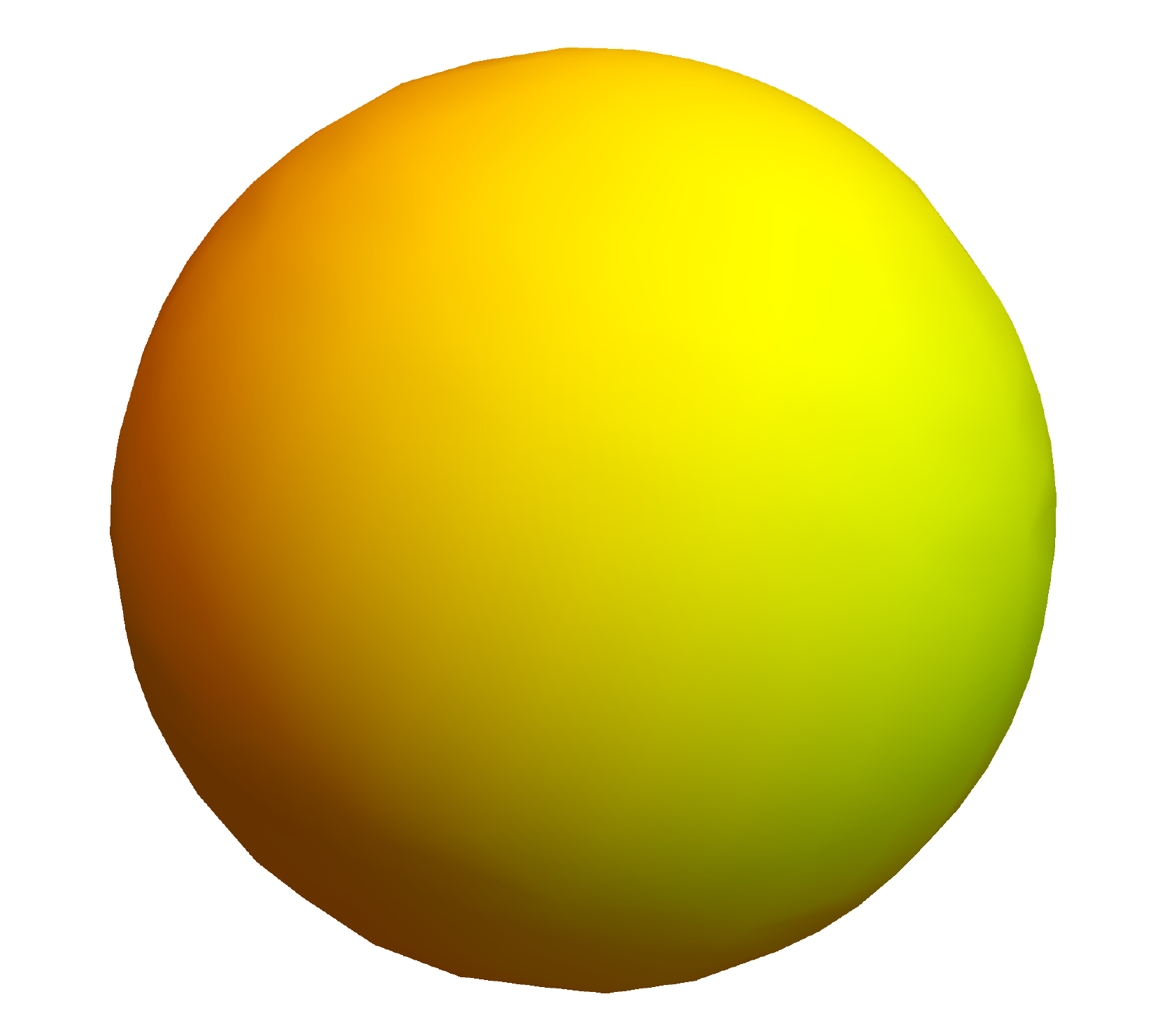}}
	\vspace{-3mm}
\caption{Sequence of the donut-like transition $f(\theta)= \epsilon P_2 (\theta) $,  for negative values of the parameter $\epsilon=0.2, 0.6, 0.8$. The waist is formed
at $\theta=\pi/2$.}
\vspace{-2mm}
\label{APP2}
\end{figure}

A different way to describe this transition is by applying pressure on the neck of the vesicle\cite{svetina} and a third one is to join two membranes along a common edge\cite{zia,jiang,yang}. 
These are three different ways of approaching the problem. In this work we will develop the first approach that has the advantage of allows control over the geometry.
Indeed, for negative values of the $\epsilon$ parameter the membrane takes a stomatocyte-like shape that can be use to model red blood cells for instance, and also can be obtained experimentally \cite{kas}. For larger negative values of the parameter, the vesicle shape tends to a donut, as shown in Fig. \ref{APP2}.

It is clear that the distribution of stress along the membrane plays an important role in these shape transitions. Following the route of the stress tensor\cite{fournier,jem,barbetta}, in section \ref{BST} we develop a theoretical framework to obtain the induced stress by harmonic deformations on spherical vesicles.
A consequence of the deformation of the spherical vesicle is the appearance of a non-trivial force along the normal direction of the membrane, so that, in order to preserve equilibrium, the pressure difference must be modified to balance these forces. As expected, the normal balance does not involve the surface tension but only bending stiffness and variations of mean curvature. As we shall see, a first integral can be obtained from this equation. Together, the normal and tangential balance equations give rise to a generalized local Young-Laplace equation given by Eq. \eqref{MGLM}, which is our principal result.

When considering these axial perturbations, the values $ l = 0, 1 $ correspond to Euclidean motions such that energy remains invariant. The first nontrivial deformation is therefore $ P_2 (\theta) $.
In addition, from the second variation of the functional \eqref{CAHEL} we know that the energy of these deformed spheres is larger than those for the spherical vesicles\cite{golushko1}. From this analysis we also known that when the pressure jump across the membrane $ P <-12 \kappa / R^3 $,  the vesicle becomes unstable with respect to the $P_2 $-deformed sphere; although has been shown that adding adhesion stabilize  membrane necks\cite{agudo}. 
Hence, after studying the geometry of the almost spherical vesicle in section \ref{QSV}, we particularize deformations by parametrizing them with axial functions $P_l(\theta)$ and calculate the induced stress for the different stages of the budding transition when $l=2$.
Due to the axial symmetry the global force just depends on the polar angle. In the north hemisphere the normal elastic force points outside the vesicle, the force vanishes at the pole and, as we approach the waist, it grows very fast and reaches a maximum value, then it decreases very fast again until it vanishes at the waist. This behavior is asymmetric with respect to the equator. If the magnitude of the deformation is small, the correction of the tangential force goes downwards as the force due to the surface tension reaches a minimum value at the waist of the peanut. However, if the magnitude of the perturbation is greater than a certain threshold value, a membrane patch appears around the waist of the peanut where the force points upward and reaches its maximum value at the waist. This was done at the end of section \ref{QSV}. In section \ref{SUM}, a summary and discussion of the obtained results is given.

\section{Bending stress tensor: local and global balance}\label{BST}

A parametrized surface embedded in $\mathbb{R}^3$ with cartesian coordinates ${\bf  x}= (x^1, x^2, x^3)$, can be  specified through the functions  ${\bf x}= {\bf X}(\xi^a)$, where $\xi^a$ $ (a=1, 2)$, are local coordinates on the surface. The infinitesimal 3D euclidean distance $ds^2=d{\bf x}\cdot d{\bf x}$ induces the corresponding arclength distance on the surface $ds^2= g_{ab}d\xi^a d\xi^b$, where $g_{ab}={\bf e}_a \cdot {\bf e}_b$ is the induced metric and  ${\bf e}_a=  \partial_a {\bf X}$ are the two local tangent vector fields. Correspondingly, the induced metric defines a covariant derivative  on the surface denoted by $\nabla_a$.
The unit normal to the surface ${\bf n}\,=  (\bar{\varepsilon}^{ab}/2) {\bf  e}_a \times{\bf e}_b $, where the symbol $\bar{\varepsilon}^{ab}= \epsilon^{ab} / \sqrt{g} $
and $ \epsilon^{ab}$ is the Levi-Civita alternating tensor and $g= \det \left(g_{ab}\right)$.
The Gauss equation, $ \nabla_a {\bf e}_b= -K_{ab}{\bf n}$, describes the change of the tangent vector fields along the surface. The extrinsic curvature 
components $K_{ab}= -\nabla_a {\bf e}_b\cdot {\bf n} $,  and  the gaussian curvature  ${\cal R}_ G$ are related through 
the Gauss-Codazzi equation, $K_{a}^cK_{cb}=KK_{cb}-g_{ab}{\cal R}_G$, and its contraction $K^{ab}K_{ab}= K^2-2{\cal R}_G$.   
The Codazzi-Mainardi equation $\nabla_a K^{ab}=\nabla^b K$ will be also useful within the surface geometry analysis\cite{willmore}.

Let us consider deformations of  the energy functional Eq. \eqref{CAHEL} under infinitesimal deformations ${\bf X} \to {\bf X} +\delta {\bf X} $ such that
\begin{equation}
\delta {\cal H}= \int dA\, {\cal E}\,  \delta{\bf X} \cdot {\bf n}  +\int dA \nabla_a Q^a, 
\end{equation}
where ${\cal E}$ is the Euler-Lagrange derivative and $Q^a$ the Noether charge. Under an infinitesimal  
translation~$\delta{\bf  X}={\bf a}$ we can write
\begin{equation}
\delta {\cal H}= {\bf a}\cdot  \int dA\, ({\cal E}\,   {\bf n} - \nabla_a  {\bf f}^a), 
\end{equation}
and therefore, as a consequence of the invariance under translations
\begin{equation}
\nabla_a {\bf f}^a= {\cal E} {\bf n}, 
\end{equation}
where the bending stress tensor  can be written as
\begin{equation}
{\bf f}^a= f^{ab} {\bf e}_b + f^a {\bf n},  
\end{equation} 
and the projections are given by
\begin{eqnarray}
f^{ab}&=& \kappa (K-K_0)\left[ K^{ab} - \frac{1}{2}(K-K_0) g^{ab} \right]  - g^{ab}\sigma, \nonumber\\
f^a &=&-\kappa \nabla^a K. \label{PRJJ}
\end{eqnarray}
For closed membranes the pressure jump $P$ can be obtained as
\begin{equation}
\nabla_a {\bf f}^a = P\, {\bf n},
\end{equation}
that after integration we have
\begin{equation}
\oint \limits_{\cal C} {\bf f}^al_a  = P\,\int \limits_{\cal M} dA\,  {\bf n}, \label{BBGG}
\end{equation}
where ${\cal C}$ is the boundary of the patch  ${\cal M}$.
On the left hand side of Eq. \eqref{BBGG} we have the elastic force on the loop, 
whereas the right hand side gives the force coming from the pressure difference $P$.
Eq. \eqref{BBGG} gives a global balance force equation along the vesicle.
 
Nevertheless we can find the local balance equations, for instance 
since the unit normal can be written as a surface divergence then we have\cite{laplace} 
\begin{equation}
{\bf n}= \frac{1}{2}\nabla_a {\bf N}^a,
\end{equation}
where ${\bf N}^a=\bar{\varepsilon}^{ab}{\bf X}\times {\bf e}_b$, a  local force balance  equation can be written as
\begin{eqnarray}
{\bf f}^a l_a &=& {\bf N}^a l_a, \nonumber\\
&=&\frac{P}{2}  {\bf X}\times {\bf T}. \label{YYLM}
\end{eqnarray}
The left hand side of Eq. \eqref{YYLM} is the elastic force 
acting on the loop  $\cal C$, with tangent $\bf T$ and binormal $\bf l$ in the Darboux frame. The right hand side corresponds to
the force due to the difference in pressure $P$.  
By writing ${\bf X}\times {\bf T}= X_l{\bf n} - X_n {\bf l}$, 
and
$
{\bf f}^al_a= F_T {\bf T}+ F_l {\bf l} + F_n{\bf n },\label{FFG}  
$
we have that
\begin{eqnarray}
F_n&=& \frac{P}{2} X_l,\nonumber\\
F_l&=& - \frac{P}{2} X_n.\label{LN}
\end{eqnarray}
where $X_l= {\bf X}\cdot {\bf l}$ and $X_n= {\bf X}\cdot {\bf n}$, 
and the projections of the force per unit length  are given by  
\begin{eqnarray}
F_n&=&-\kappa \nabla_l K,\label{PROJ}\\
F_l &=&-\Sigma+   \frac{\kappa}{2}\left( K_T^2 -K_l^2 +2 K_0K_l \right). \label{POJ}
\end{eqnarray}
Where we have introduced the following notation $K_l= K_{ab}l^al^b$, $K_T=K_{ab}T^aT^b$,
$K_\tau= K_{ab}l^aT^b $ and $\nabla_l K =l^a\nabla_a K$.
The first  equation in \eqref{LN} describes the local force balance along the normal direction to the surface, on the sphere $X_l= R\,  {\bf n}\, \cdot ({\bf T}\times {\bf n})=0$. 
The second equation in  \eqref{LN} instead describes the balance force along the binormal direction, for instance
on the sphere $F_l=-(\sigma + \kappa K_0^2/2 ) + \kappa K_0/R$ and $ X_n= R\,  {\bf n}\cdot {\bf n}=R$,  where $R$ is the radius of the sphere, and thus it reduces to the classical Young-Laplace equation\cite{lubarda}.
Since in both equations \eqref{LN} the pressure jump appears, we add them to obtain the following relation
\begin{equation}
\Sigma - \frac{\kappa}{2}\left( K_T^2 -K_l^2 +2 K_0K_l \right)- \kappa \nabla_l K=\frac{P}{2}\left( X_n + X_l \right)\label{MGLM},
\end{equation}
where $\Sigma$ was defined on Eq. \eqref{LPU}.
It is interesting that, as far as we know,  Eq. \eqref{MGLM}  had not been reported before, it comes from the local force balance Eqs. \eqref{LN} and it is the first main result of this paper.

\section{Quasi-spherical vesicles}\label{QSV}

In order to obtain the Monge gauge around a sphere of radius $R$, let us 
parametrize an almost spherical surface as follows
\begin{equation}\label{qsp}
{\bf X}(\theta, \phi) = (r\sin\theta \cos\phi, r\sin\theta \sin\phi, r\cos\theta),
\end{equation}
where $ r(\theta, \phi)= R\left[1 +  \epsilon f(\theta, \phi)\right]$ and $\epsilon \lesssim1$ is the deformation parameter. Without loss of generality let 
us study the case  $ f=f(\theta)$, that corresponds to perturbations with cylindrical symmetry. 
We write the tangent vector fields to the surface and the unit normal as
\begin{eqnarray}
{\bf e}_\theta &=& r\boldsymbol{\theta} + r' {\bf r}, \nonumber\\
{\bf e}_\phi &=& r\sin\theta\,  \boldsymbol{\phi}, \nonumber\\
{\bf n}&=& \frac{1 }{ \sqrt{r^2 +r'^2 }} ( r\, {\bf r}  - r' \boldsymbol{\theta}), 
\end{eqnarray}
where $'$ denotes derivative respecto to $\theta$ and the unit basis is the following as usual
\begin{eqnarray}
{\bf r}&=& ( \sin\theta\cos\phi , \sin\theta\sin\phi , \cos\theta )\nonumber\\
\boldsymbol{\theta}&=& (\cos\theta\cos\phi , \cos\theta\sin\phi , -\sin\theta ) \nonumber\\
\boldsymbol{\phi}&=&( -\sin\phi , \cos\phi  , 0  ).
\end{eqnarray}
Thus, the components of the induced metric can be written as
\begin{equation}
g_{ab} d\xi^a d\xi^b = (r^2 + r'^2)d\theta^2 + r^2\sin^2\theta d\phi^2. 
\end{equation}
The determinant  of the induced metric  is therefore
\begin{equation}
g=(r^2 + r'^2)r^2 \sin^2\theta, 
\end{equation}
so that the area element $dA= \sqrt{ r^2 + r'^2} \, r   \sin\theta \,d\theta \, d\phi$.
The extrinsic curvature components are given by
\begin{eqnarray}
K_{\theta \theta} &=&  \sqrt{r^2 +r'^2} - r \partial_\theta \left(  \frac{ r' }{  \sqrt{r^2 +r'^2 }} \right) \nonumber\\
&&+ r' \partial_\theta \left( \frac{r}{ \sqrt{r^2 +r'^2 } } \right),  \nonumber\\
K_{\phi\phi} &=&  \frac{1 }{ \sqrt{r^2 +r'^2 }} \left(   r^2\sin^2\theta- rr' \sin\theta\cos\theta \right), \nonumber\\
K_{\theta\phi}&=&0, 
\end{eqnarray}
while the mean curvature can also be obtained as
\begin{eqnarray}
K &=&\frac{K_{\theta\theta }}{ r^2+ r'^2} + \frac{K_{\phi\phi}}{r^2\sin^2\theta }, \nonumber\\
&=& \frac{2}{  { \sqrt{r^2 +r'^2 } } }  - \frac{r}{ r^2 +r'^2 }\partial_\theta \left(  \frac{ r' }{  \sqrt{r^2 +r'^2 }} \right) \nonumber\\
&&+\frac{r'}{r^2+r'^2 } \partial_\theta \left(  \frac{ r }{  \sqrt{r^2 +r'^2 }} \right) - \frac{r'\cot\theta}{r \sqrt{r^2+r'^2} }. 
\end{eqnarray}
The gaussian curvature can be obtained in several different ways. By using derivatives of the unit normal 
we have that 
\begin{eqnarray}
{\cal R}_G&=& g^{\theta\theta} g^{\phi\phi } {\bf n}\cdot (\partial_\theta {\bf n} \times \partial_\phi {\bf n}) \nonumber\\ 
&=&\frac{r^2}{ (r^2+r'^2)^2}  - \frac{r}{ (r^2+r'^2)^{3/2} }\partial_\theta\left(  \frac{r'}{ \sqrt{r^2+r'^2} }    \right)\nonumber\\
&&\times \left(1- \frac{r'\cot\theta}{r }  \right).
\end{eqnarray}
As shown in the appendix, the corresponding expressions with $ f (\theta, \phi) $ can also be obtained straightforwardly. As for the flat case, only expansions up to order $\epsilon$ wil be considered in these formulas.
\begin{figure}[htp!]  
\centering 
\subfigure[]{
\includegraphics[scale=0.24]{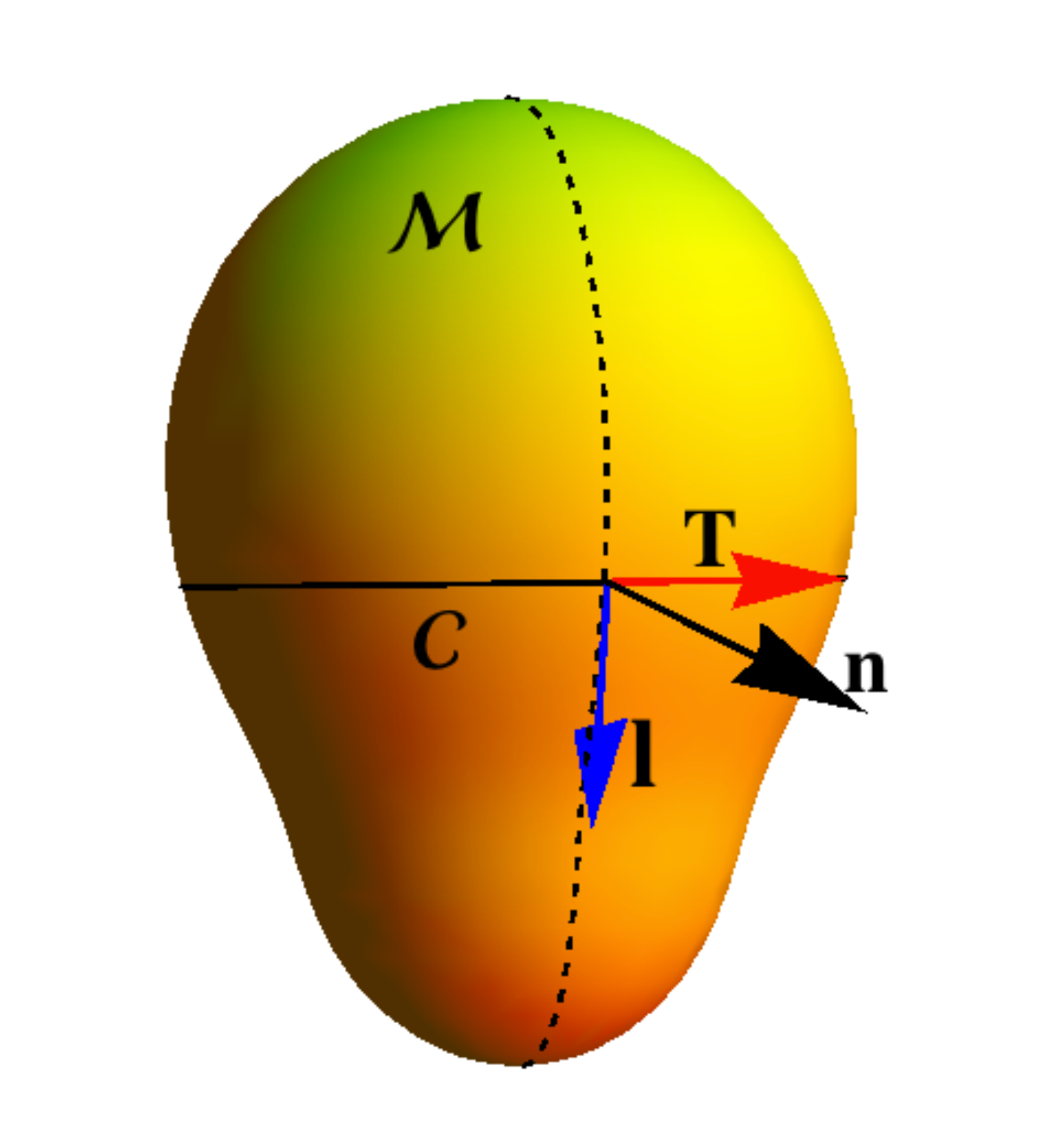}}
\hspace{0.5mm}
\subfigure[]{
\includegraphics[scale=0.25]{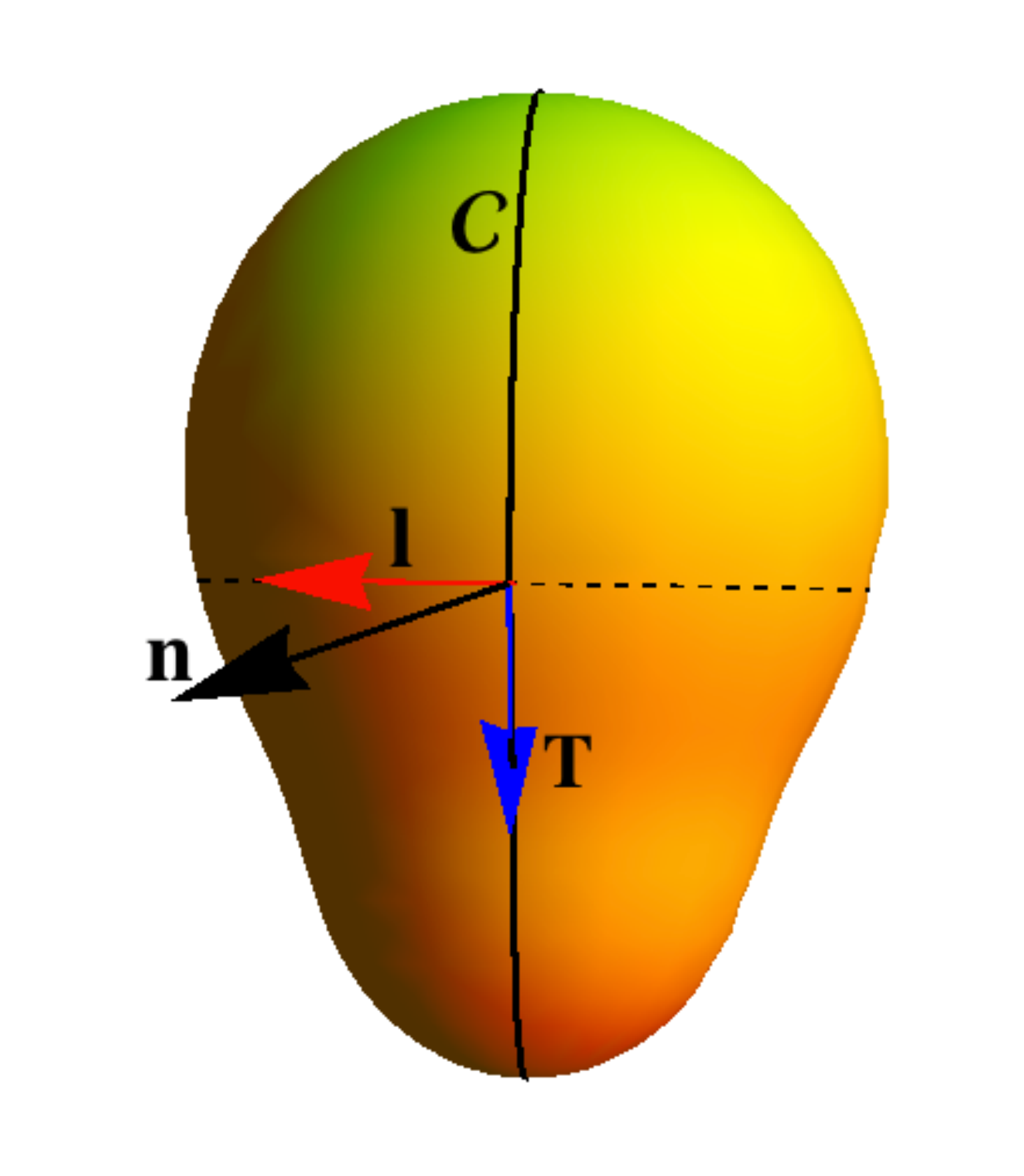}}
\caption{ The Darboux frame adapted to the curve $\cal C$  onto a deformed sphere:  a) $\cal C$ as a  parallel, b) $\cal C$ as a meridian.  
The unit tangent $\bf T$, the unit normal $\bf n$, and the binormal vector ${\bf l}= {\bf T}\times {\bf n} $. }\label{NNCC}
\vspace{-2mm}
\end{figure}

\subsection{Local balance}

Taking advantage of the axial symmetry of the problem, let us select the curve $\cal C$ as a parallel 
such that $\theta=\theta_0$, see for instance Fig. \ref{NNCC}a.  
The  unit normal  to ${\cal C}$, tangent to the surface is given by 
\begin{equation}
{\bf l}= \frac{1}{ \sqrt{r^2+r'^2 }} (r \boldsymbol{\theta} + r' {\bf r} ),
\end{equation}
and thus we see that  $l_\theta=\sqrt{r^2+ r'^2 }  $ and $l_\phi=0 $ 
so that along this loop we have 
\begin{eqnarray}
K_l &=& K_{ab}l^a l^b, \nonumber\\
&=&\frac{  K_{\theta\theta} }{ r^2 + r'^2 }, \nonumber\\
&=& \frac{1}{\sqrt{r^2+r'^2}}\left(1+ \frac{r'^2+ rr''}{r^2+r'^2}  \right). 
\end{eqnarray}
Notice that $K_l$ is  the normal curvature of a meridian on the  deformed sphere, see Fig. \ref{NNCC}(a).
We can see that  ${\bf T}=\boldsymbol{\phi}$ and therefore we have $T_\theta=0$ and $T_\phi=r\sin\theta$ 
so that
\begin{eqnarray}
K_T&=& \frac{ K_{\phi\phi}} { r^2\sin^2\theta}, \nonumber\\
&=& \frac{1}{\sqrt{r^2+r'^2} }\left(1- \frac{r'\cot\theta}{r}  \right),
\end{eqnarray}
its the normal curvature of ${\cal C }$ itself.
We also see that  the gaussian torsion of ${\cal C} $ vanishes, i.e. $ K_\tau= 0$, as a consequence of the axial symmetry.
Projections of the force can then be obtained in the following way
\begin{eqnarray}
F_T&=& 0,   \nonumber\\
F_n&=& -\frac{\kappa K'}{ \sqrt{r^2+ r'^2}}, 
\end{eqnarray}
According to Eq. \eqref{POJ},  where $F_l$ was given, if the spontaneous curvature is zero, the term corresponding to bending force in $F_l$ points downward, i. e. in the opposite direction to the surface tension component, at points where the normal curvature of the loop $ \cal C $ is greater than the normal curvature of the meridian i. e., $ K_l ^ 2 <K_T ^ 2 $. Such component points upwards when $K_T^2 < K_l^2$. Furthermore, this component of the force vanishes at umbilical points $\theta=\theta_c$ where  $K_l^2=K_T^2$. Clearly, on the unperturbed sphere, the bending force vanishes identically.

\begin{figure}[h!]
\centering 
	\includegraphics[width=2.15in, angle=-90]{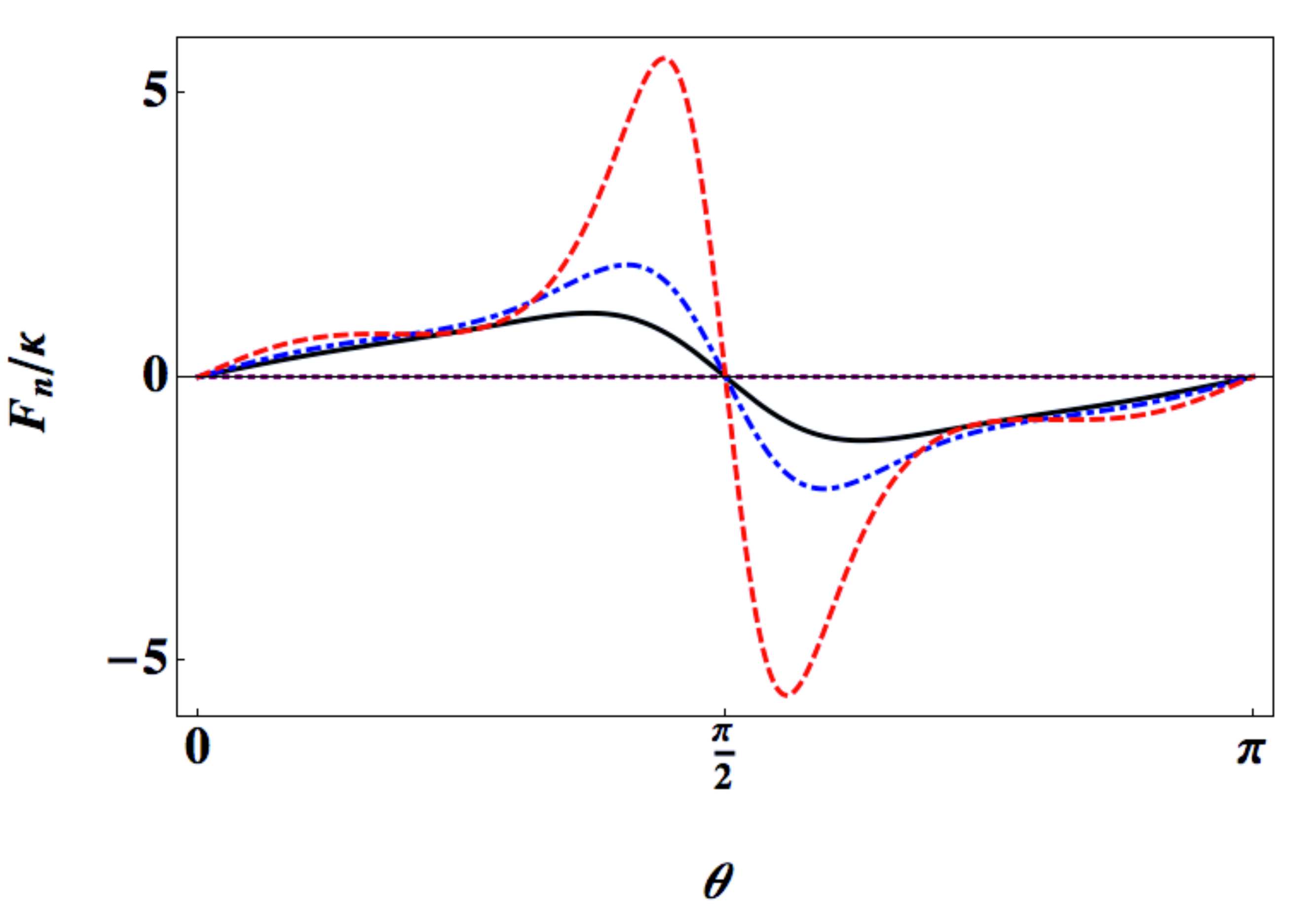}
\hspace{- 0.5mm}
\caption{Normal projection of the force $F_n/\kappa$ for some values of deformation parameter. For $\epsilon=0$ (dashed horizontal line), $\epsilon=0.2$ (continuos line), $\epsilon= 0.3$ (dot-dashed line), $\epsilon=0.5$ (dashed line). If  $\epsilon$ is positive force goes outward along the unit normal of the peanut shape.}
\label{TN}
\end{figure}

In the case when spontaneous curvature is non-zero, it contributes nontrivially to this force. For example, in the budding transition shown in Fig.  \ref{APP}, the normal force $F_n$ has been plotted in Fig.\ref{TN}. At the pole the force is zero, then increases to a maximum that is reached at a value close to $\theta=\pi/2$, then the force decreases until vanishes at the waist of the peanut shape.

\begin{figure}[h!]
\centering 
\hskip -5mm
	\includegraphics[width=2.15in, angle=-90]{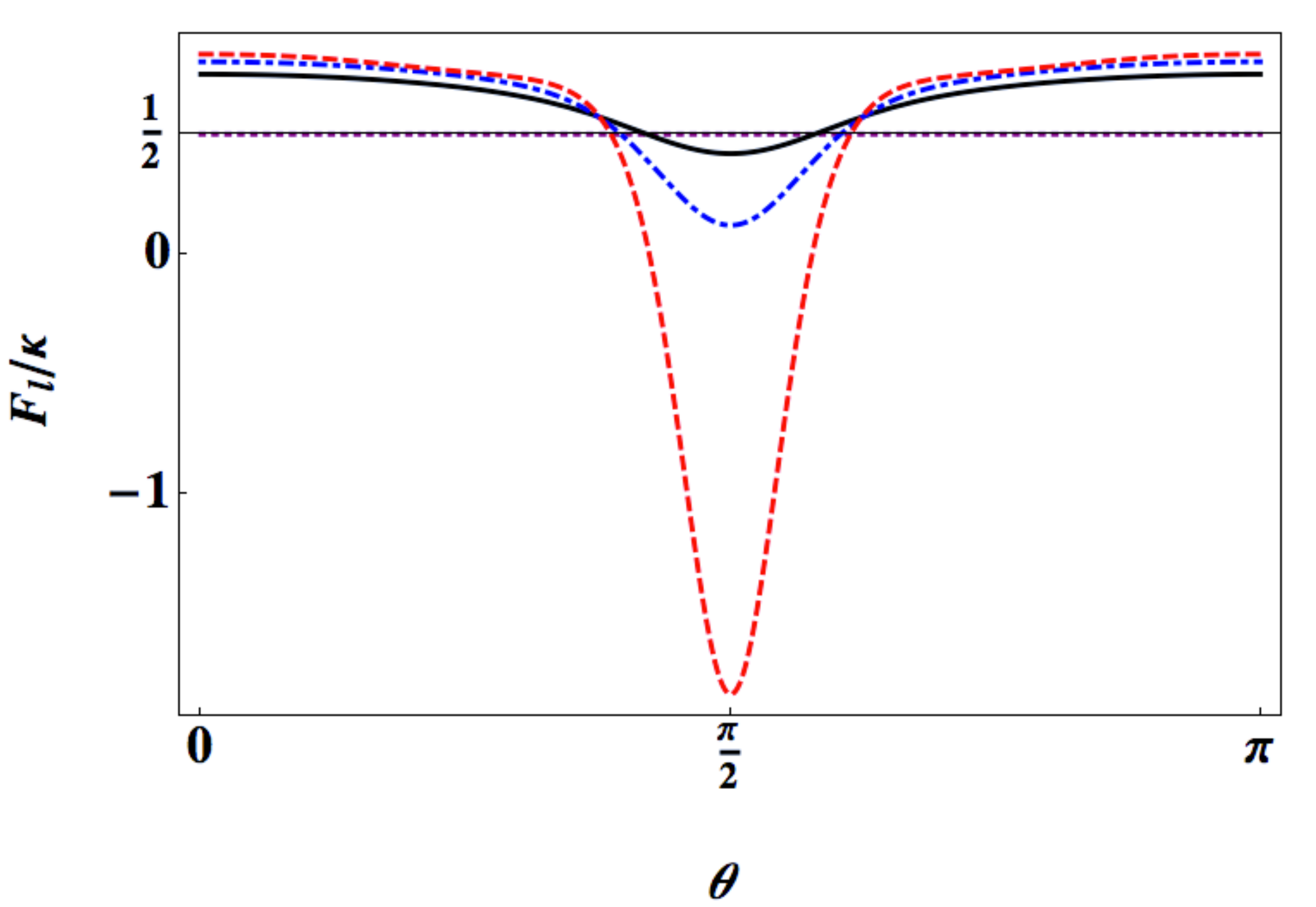}
\hspace{-1mm}
\caption{Binormal projection of the force $F_l/\kappa$ for some values of deformation parameter.  For $\epsilon=0$ (dashed horizontal line), $\epsilon=0.2$ (continuos line), $\epsilon= 0.3$ (dot-dashed line), $\epsilon=0.5$(dashed line). If  $\epsilon$ positive the force goes downward along the binormal. Note that
in this last configuration there is a neighborhood around $\theta=\pi /2$ where the force goes upward.}
\label{TAN}
\end{figure}

The tangential force is shown in Fig. \ref{TAN} for the case $K_0=1$. To first order in the parameter $\epsilon$, this force points downward along the binormal direction ${\bf l}$, see Fig. \ref{NNCC}a, reaching its lowest value at the waist. For larger values of $\epsilon$, a region appears where the force becomes negative and points upward, reaching its highest value at the waist.

Let us make a more detailed analysis of the balance equations up to first order.
For parameterization \eqref{qsp} we have that $X_T= {\bf X}\cdot {\bf T}=0$, $X_n= r^2/\sqrt{r^2 + r'^2}$ and $X_l= rr'/ \sqrt{r^2+r'^2}$,
and therefore the local balance given in Eqs. \eqref{LN} turns respectively into
\begin{eqnarray}
&& -\kappa \frac{ (K'_T + K'_l)}{\sqrt{r^2+r'^2}} = \frac{P}{2} \frac{ rr'}{\sqrt{r^2+r'^2} }, \label{BANCE}\\
&& \Sigma  - \frac{\kappa}{2}\left( K_T^2 -K_l^2 +2 K_0K_l \right)= \frac{P}{2} \frac{ r^2}{\sqrt{r^2 + r'^2} } \label{BALANCE}
\end{eqnarray}
We realize that the normal balance Eq. \eqref{BANCE} does not involve the surface tension $\sigma$ but only
the rigidity $\kappa$ and then it can be rewritten as
\begin{equation}
-\kappa  K'= \frac{P}{4}  \left(r^2  \right)' ,
\end{equation}
which implies that
$
-\kappa  K = \frac{P}{4} r^2 + C, \label{BMW}
$
where $C$ a constant that can be determined by taking the unperturbed sphere $r=R$, such that~$K=2/R$ and
\begin{equation}
C=- \left( \frac{2\kappa}{R} + \frac{PR^2}{4}\right).
\end{equation}
This implies that the mean curvature of the deformed sphere can be expressed as modifications, due to the pressure and of the 
bending stiffness, of the spherical case as follows
\begin{equation}
K= \frac{2}{R} + \frac{PR^2}{4\kappa} \left( 1- \frac{ r^2 }{R^2} \right).\label{KLP} 
\end{equation}
In the region where $r^2/R^2 <1$ i.e., at points where the deformation function $ f < 0$, the mean curvature becomes greater that the spherical case $K\sim 2/R +  PR^2 / (4\kappa)$. In the same way when $r^2/R^2 >1$,~($f>0$) then, the curvature is smaller through $K\sim 2/R - Pr^2/ (4\kappa )$. At points where $f=0$ we regain the spherical case locally $K=2/R$. 
It is worth mentioning some specific data, for lipid  membranes the bending constant typically\cite{galatola} is $\kappa\sim 80 \times 10^{-21} J$, whereas
$R\sim 10^{-6}m $. For these kind of vesicles the correction term $ PR^3 /(8\kappa)$ will be $50\%$ relevant if $P\sim 3 \text{Pa}$.

\begin{figure}[h!]
\centering 
	\includegraphics[width=3.6in]{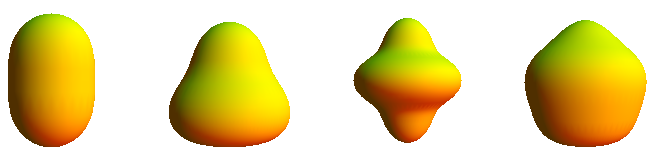}
\hspace{- 1mm}
\caption{Deformed spheres with $P_l(\theta)$ for modes  $l=2, 3, 4,  5$ respectively.}
\label{PPE}
\end{figure}

The mean curvature Eq. \eqref{KLP} can also give rise to the following expression
\begin{equation}
\frac{2}{R}+ \frac{PR^2}{4\kappa} \left( 1-\frac{r^2}{R^2} \right)= \frac{1}{\sqrt{r^2+r'^2}} \Big( 2+ \frac{r'^2 + rr''}{r^2+r'^2 }
- \frac{r'}{r}\cot\theta \Big).\label{BNM} 
\end{equation}
In a general setting,  the Lagrange multiplier $P$ must be substituted into Eq. \eqref{BALANCE}  and then solve for the variable $r$, what will determine the multiplier $\sigma$, i.e. solve the shape equation.
Instead of that, we  propose solutions in terms of the Legendre polynomials $P_l(\theta)$, that satisfies the Legendre
equation with azimuthal symmetry, $ f'' + \cot\theta f' + l(l+1)f=0$.  So, up to first order Eq. \eqref{BNM} turns into
\begin{equation}
f''- \cot\theta f' - \beta f=0, \label{LLM}
\end{equation}
where $\beta= PR^3/ (2\kappa)+2$.
The mode $P_l(x)$ with $x=\cos\theta$,  becomes compatible with the normal force balance \eqref{LLM}, if the relation
$3x \dot P_l (x) = [l(l+1)+ \beta] P_l(x)$ is fulfilled. That is if $\beta\, \sin^2\theta =  l(l-2) \cos^2\theta - l(l+1).$
It determines the pressure difference $P$ up to first order.  
Take for instance the first non-trivial deformation $l=2$, see Fig. \ref{PPE}, for which the 
pressure difference $P$ is found to be 
\begin{equation}
P=-\frac{4\kappa}{ R^3}\left( 1+ 3\csc^2\theta   \right).
\end{equation}
The lowest value of $P$  is reached at the waist of the vesicle $\theta=\pi/2$. Beyond this point the pressure increases and becomes very large near the poles.
Once the pressure jump $P$, the multiplier that fixes the volume inside the vesicle, has been determined, the value of $\sigma$ will be fixed by the balance equation along the binormal, Eq. \eqref{BALANCE}.

The relation with the surface tension $\sigma$ is given by  balance equation along the binormal Eq. \eqref{BALANCE}, so up to first order Eq. \eqref{MGLM} can be written as 
\begin{equation}
\Sigma -\frac{\kappa K_0 }{R}  + {\cal F}\epsilon
=\frac{PR}{2}[1+ (f+f')\epsilon],
\end{equation}
where we introduced the correction function 
\begin{eqnarray}
{\cal F}&=&\frac{\kappa}{R^2} \Big[  \cot\theta f' - f'' + RK_0(f+ f'' ) \nonumber\\
&&+ (1- \cot^2\theta)f' + \cot\theta f'' + f''' \Big],\label{corrF0}
\end{eqnarray}
that can in turn be rewritten, with use of Legendre equation, as
\begin{eqnarray}
{\cal F} &=& \frac{\kappa}{R^2} [ (2 - RK_0) \cot\theta +2 -l(l+1)  ]f' \nonumber\\
&&+ \frac{\kappa}{R^2}[ (1- RK_0) l(l+1) -RK_0  ]f.\label{corrF}
\end{eqnarray}
Therefore, up to first order in the deformation parameter $\epsilon$, we can write the local Young-Laplace law as follows
\begin{eqnarray}
\frac{PR}{2}= \left( \Sigma - \frac{\kappa K_0}{R} \right)[1 - (f+ f')\epsilon  ] + {\cal F}\epsilon.  \label{YOLA} 
\end{eqnarray}
The case with $f=0$ reproduces exactly Eq. \eqref{LPU}, where the pressure difference 
is a constant. 

\begin{figure}[th]
\centering \hskip -5mm
	\includegraphics[width=2in, angle=-90]{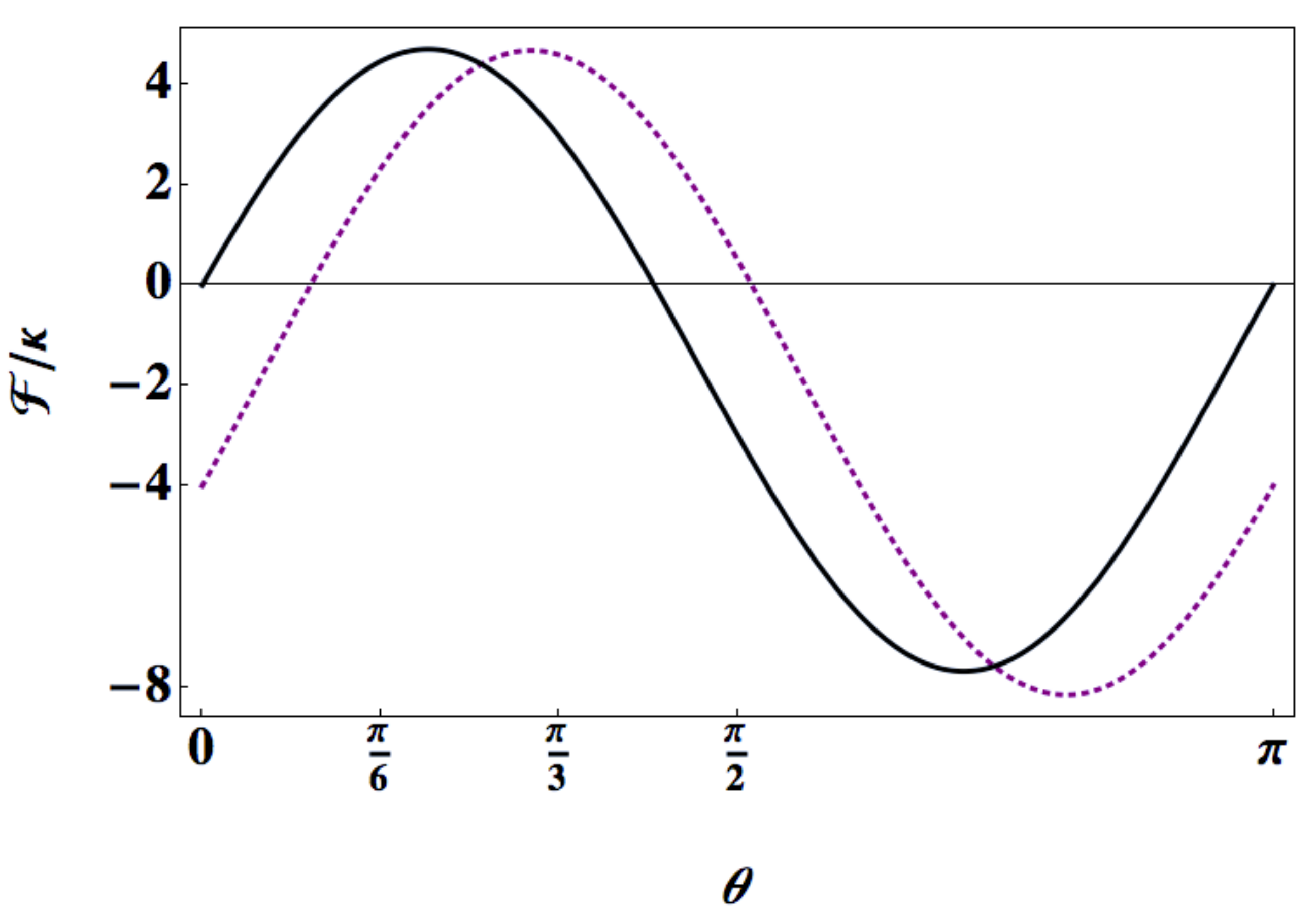}
\hspace{- 0.5mm}
\caption{Correction function $\cal F/ \kappa$ Eq. (\ref{corrF}) for a unit deformed sphere with $P_2(\theta)$. Continuos line 
corresponds to $K_0=0$, the dashed line is for $K_0=1$.}
\label{FL2}
\end{figure}

Is worth  noting that Eq.\eqref{YOLA} is valid for any perturbation $f=P_l(\theta)$, where $l=2, 3, \dots$ As far as we know, this is the first time that this equation is obtained. Take for instance the case where $f=P_2( \theta)$. The correction function associated to the bending ${\cal F}(\theta)/\kappa$, have been plotted in Fig. \ref{FL2}.
Dashed line corresponds to the case when spontaneous curvature $K_0=1$, and the continuos line to $K_0=0$.
Although their shape is very similar with positive values and a maximum in the northern hemisphere, observe that whereas the correction vanishes at the poles if $K_0=0$, there is a negative correction there and vanishes at the 
waist if $K_0=1$.

\subsection{Global forces}

A complementary global  analysis can be done starting with the left hand side of Eq. \eqref{BBGG} that is the total force on a closed horizontal loop $\cal C$- The integral is made from an initial angle $\theta_0$
\begin{eqnarray}
{\bf F} (\theta_0) &=&\oint \limits_{\cal C} ds\,  {\bf f}^a l_a, \nonumber\\
&=&  \frac{ 2\pi r\sin\theta_0 }{ \sqrt{r^2+r'^2}  }  \Big[  F_n (r\cos\theta_0 -r'\sin\theta_0 )  \nonumber\\
&&+ F_l ( r'\cos\theta_0 - r\sin\theta_0) \Big] \, {\bf k}.\label{FG1}
\end{eqnarray}
That can be rewritten as 
\begin{equation}
{\bf F}(\theta_0)= [ F_1(\theta_0) +  F_2(\theta_0)]{\bf k}, 
\end{equation}
where we have defined the functions
\begin{eqnarray}
&&F_1(\theta_0)= \left( \frac{2\pi r \sin\theta_0}{ \sqrt{r^2 + r'^2 } }   \right)( r\sin\theta_0 -r' \cos\theta_0 )
\left( \sigma + \frac{\kappa K_0^2}{2}\right),  \nonumber\\
&& F_2(\theta_0) = \frac{ 2\pi \kappa\,   r\sin\theta_0 }{ \sqrt{r^2+r'^2}  }  
\Big[\frac{K'}{\sqrt{r^2+r'^2}} (r'\sin\theta_0-r\cos\theta_0 )  \nonumber\\
&&+ \frac{1}{2}\left( K_T^2 -K_l^2 +2 K_0K_l \right)( r'\cos\theta_0 - r\sin\theta_0) \Big].
\end{eqnarray}
On the sphere and with zero spontaneous curvature $r=R$,  and thus the force comes just from the surface tension i.e., 
$ F_1(\theta_0)= 2\pi \sigma R \sin^2 \theta_0$ and 
$F_2(\theta_0)=0$. In equilibrium this force must be balanced with the  Laplace pressure given by $2\pi r^2 \sin^2\theta_0 P/2$ in such a way that Young-Laplace law emerges.
On the sphere with non-zero spontaneous curvature, we have that
 $F_1(\theta_0)=2\pi R \sin^2 \theta_0 (\sigma + \kappa K_0^2/2) $
 and $F_2(\theta_0)= -2\pi \kappa  K_0 \sin^2 \theta_0 $, such that
 \begin{equation}
{\bf F}(\theta_0)=  2\pi R \sin^2 \theta_0 \left[ \sigma + \frac{ \kappa K_0}{2} \left(K_0- \frac{2}{R}   \right)  \right] {\bf k}.
\end{equation}
In this case the spontaneous curvature determines the behavior of the force. If $ K_0>2/R$ the correction to Laplace force becomes positive and points upward. If $0<K_0<2/R$ the correction force points downward. For $K_0<0$ the correction becomes positive. When this force is balanced with the Laplace pressure Eq. \eqref{LPU} is recovered. Therefore due to axial symmetry, both local and global balance equations give rise to the same Laplace-Young relation.

\begin{figure}[h!]
\centering \hskip -6mm
	\includegraphics[width=2.15in, angle=-90]{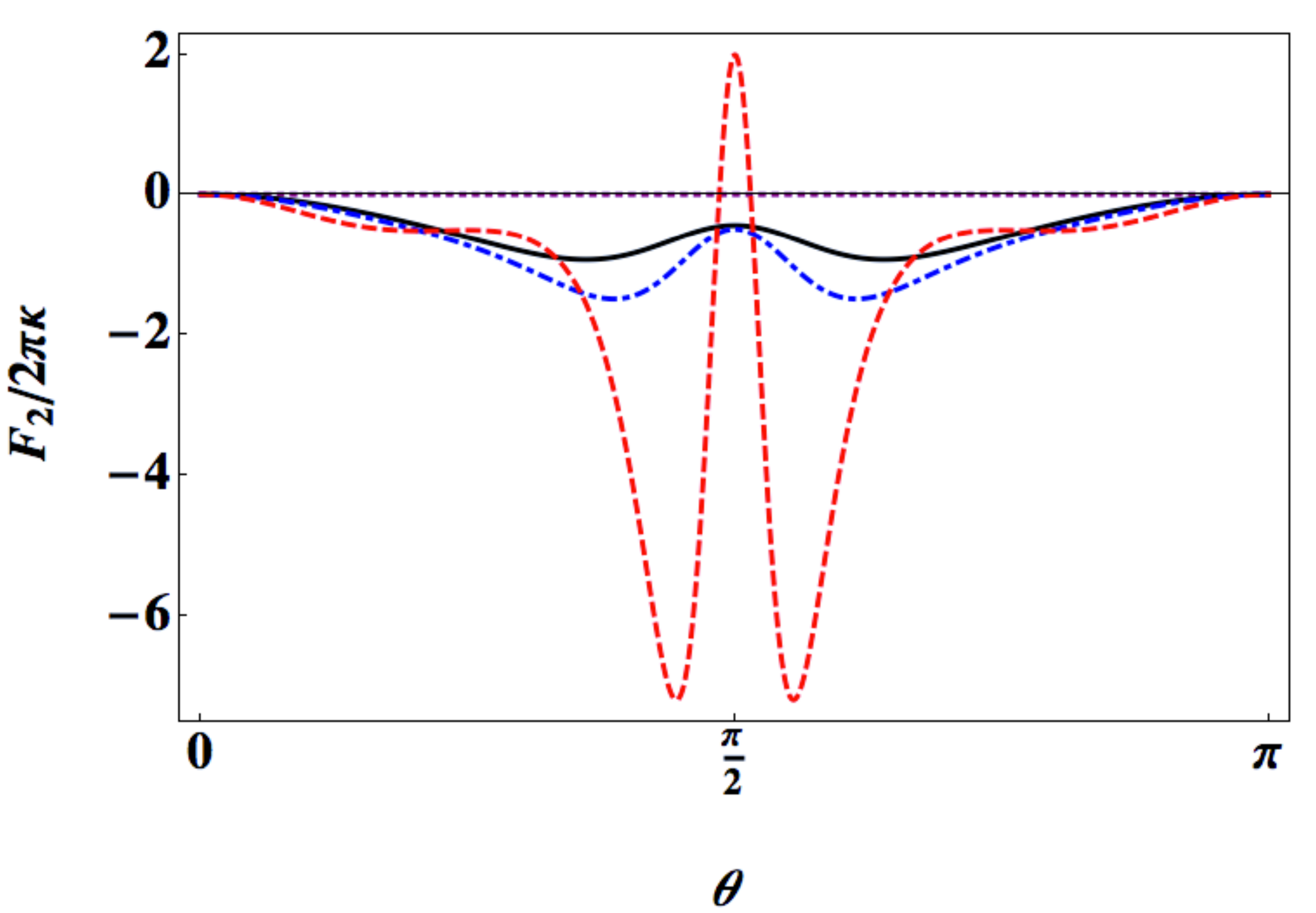}
\hspace{- 0.5mm}
\caption{Force component ${F}_2/ (2\pi \kappa)$ for a unit axially deformed sphere $P_2(\theta)$ for $K_0=0$, for some values of $\epsilon$. For $\epsilon=0$ (horizontal dashed line), 
$\epsilon=0.2$ (continuos line), $\epsilon= 0.3$ (dotdashed), $\epsilon= 0.7$ (dashed). For large values of deformation parameter there is a important behavior at the waist. 
}
\label{F22k}
\end{figure}

Figure \ref{F22k} shows the component of the force that only depends on bending and spontaneous curvature in the case of $f=P_2(\theta)$ and $K_0=0$ for some values of $\epsilon$. If the fluctuation is small the force is also small, it is negative and reaches a local maximum at the waist. However, if the fluctuation is greater than a certain threshold value, a region around the waist appears where this force becomes positive with large magnitude at the waist.

\begin{figure}[th]
\centering \hskip -6mm
	\includegraphics[width=2.15in, angle=-90]{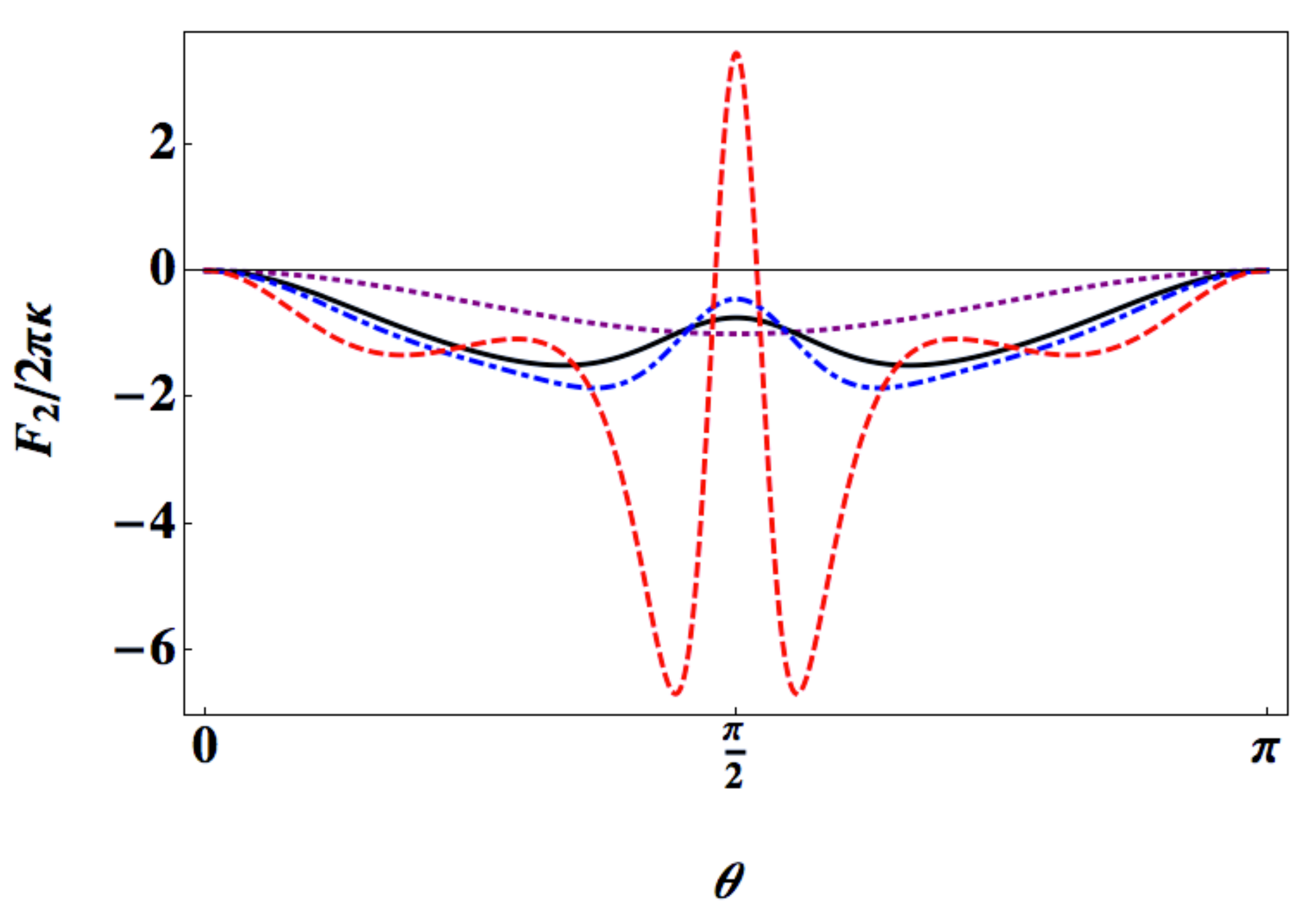}
\hspace{- 0.5mm}
\caption{Force component ${F}_2/ (2\pi \kappa)$ for a unit axially deformed sphere $P_2(\theta)$ with non vanishing spontaneous curvature $K_0=1$, for different values of $\epsilon$. For $\epsilon=0$ (horizontal dashed line), $\epsilon=0.2$ (continuos line), $\epsilon= 0.3$ (dotdashed), $\epsilon= 0.7$ (dashed). Although the behavior is similar to the previous case, the effect of spontaneous curvature is to decrease $F_2$ for small values of $\epsilon$ and increase the effect at the waist for large values of the deformation parameter.
}
\label{F33k}
\end{figure}

Figure \ref{F33k} depicts the case $K_0=1$ for some values of $\epsilon$. Although the behavior of the force $F_2$ is similar to the case with $K_0=0$, spontaneous curvature induces a deformation to the sphere, so the force for $\epsilon=0$ is non-zero and changes a bit with the angle. In addition and precisely for this reason, the force at the waist for large values of $\epsilon$ is larger than in the previous case.

To study the total force on a horizontal loop we expand the expressions for $\epsilon$ small enough, up to first order we have

\begin{equation}
F_1(\theta_0)= 2\pi R \Big[ (1+\epsilon f)\sin^2\theta_0 - \frac{\epsilon}{2} f' \sin 2\theta_0 \Big]\left( \sigma + \frac{\kappa K_0^2}{2}\right).
\end{equation}
When evaluated at the waist the force $F_1$ is given by
\begin{equation}
F_1(\pi/ 2)= 2\pi R \left(1-\frac{\epsilon P_l(\pi/2) }{2} \right) \left( \sigma + \frac{\kappa K_0^2}{2}\right).
\end{equation}
On the other hand for $F_2$ we obtain
\begin{eqnarray}
\frac{ F_2(\theta_0)}{2\pi\kappa\sin\theta_0 } &=&     
 \frac{1}{R}(f''' + f'' \cot\theta - \cos 2\theta \csc^2 \theta f')\epsilon    \cos\theta_0   \nonumber\\
 &&[ K_0 f' \epsilon \cos\theta_0 - K_0 \sin\theta_0 +  K_0 f'' \epsilon\sin\theta_0 \nonumber\\
&& - \frac{1}{R}(f'' -\cot\theta_0 f'  )\epsilon \sin\theta_0  ] \nonumber\\  
&&= \Big[  \left(\frac{1}{R} -K_0 \right)l(l+1)\sin\theta \, f \nonumber\\
&&+ (4-l(l+1))\frac{\cos\theta}{R}f'     \Big]\epsilon - K_0 \sin\theta_0,
\end{eqnarray}
that can then be written as follows
\begin{eqnarray}
F_2(\theta_0) &=& - 2\pi \kappa K_0 \sin^2\theta_0 \nonumber\\
&&+2\pi\kappa \Big[  \left(\frac{1}{R} -K_0 \right)l(l+1)\sin^2\theta \, f \nonumber\\
&&+ (4-l(l+1))\frac{\sin 2\theta}{2R}f'     \Big]\epsilon .
\end{eqnarray}
When $\theta_0=\pi/2$ we obtain the force $F_2$ on the waist, 
\begin{equation}
F_2 \left(\pi/2  \right)= - 2\pi \kappa K_0 +2\pi\kappa\epsilon \left(\frac{1}{R} -K_0 \right)l(l+1) P_l(\pi/2). 
\end{equation}
Amusingly, the second term in previous expression vanishes if  $l$ is odd since $P_l(\pi/2)=0$. Then, for asymmetric deformed spheres as those shown in Fig. \ref{PPE}, there is no such contribution to the net force.

\section{Summary and Discussion}\label{SUM}
In this work we have introduced a theoretical framework to analyze the stress induced by small deformations in spherical membranes.
The analysis we have made is based on the geometric formalism of the stress tensor and for this we have taken explicitly axially symmetric deformations parametrized by Legendre's harmonic functions $ P_l (\cos\theta)$, with only dependence 
on the  polar angle $\theta$. With this we have two main results; in the first we found conditions of local equilibrium for closed vesicles. For  small deformations, we find that the pressure difference along the membrane is no longer constant 
(to which is reduced for spherical membranes), but a correction term appears that depends on the polar angle and the $l$-mode  of the fluctuation. This can be called a local Young-Laplace equation.
In particular, figure \ref{FL2} shows the correction term for the first non-trivial mode $ l = 2 $, which describes a kind of 
budding transition.\\
As a second relevant result, we have calculated the total force on horizontal loops of the vesicle including the force due to the pressure difference $P$. Clearly, this force also depends on the $l$-mode of the fluctuation and the polar angle in addition to the bending stiffness. The equilibrium condition gives us the corresponding global Young-Laplace law as a result.

Internal molecular orientations of the membrane give rise to textures with topological defects\cite{gil}. Indeed, if the membrane has a spherical shape the topological charge must be 2. These defects and the texture itself induce stresses that must also be taken into account in a more complete analysis of closed surfaces. That subject is in progress and will be discussed later elsewhere.

The order of a nematic liquid crystal on curved surfaces is undetermined when topological defects are present, so that there is a geometric coupling between the nematic director and the shape of the surface. In addition to this, when we are in the presence of active matter that has the capacity to generate forces, the defects can be modified on the surface and indeed can move. Certainly, the geometric properties help to control the collective behavior of the active matter\cite{giomi-am1,giomi-am2,alaimo}. Therefore, the study of lipid membranes is currently of fundamental interest.

\section*{Acknowledgements}


GTV would like to thank CONACyT for support through a scholar fellowship (Grant No 381047).

\appendix 

\section{Extrinsic curvature}
The unit normal ${\bf n}$  can be obtained from its definition
$
{\bf e}_\theta \times {\bf e}_\phi = {\bar{ \epsilon}}_{\theta\phi} {\bf n},
$
where the Levi-Civita tensor ${\bar{ \epsilon}}_{\theta\phi}=\sqrt{g}\, {\epsilon}_{\theta\phi}= \sqrt{g}. $ Therefore,
\begin{eqnarray}
{\bf e}_\theta \times {\bf e}_\phi &=& (r \boldsymbol{\theta} + r' {\bf r} ) \times ( r\sin\theta \boldsymbol{\phi} ), \nonumber\\
&=&r^2 \sin\theta\, {\bf r} - r r' \sin\theta\, \boldsymbol{\theta},
\end{eqnarray}
so that
$
|| {\bf e}_\theta \times {\bf e}_\phi || = r\sin\theta \sqrt{r^2+r'^2 },
$
as was written above in the main text. However, we can also see that the normal vector could be expressed as a surface divergence. Let us define
$
 {\bf N}^a= {\bar{ \epsilon}}^{ab}{\bf X}\times {\bf e}_b, 
$
such that
\begin{equation}
\nabla_a {\bf N}^a \, =\varepsilon^{ab} {\bf e}_a \times {\bf e}_b 
\, = 2{\bf n}.
\end{equation}
Therefore, when we integrate it we have
\begin{eqnarray}
\int dA\,  {\bf n} &=& \frac{1}{2}\int dA \, \nabla_a {\bf N}^a, \nonumber\\
&=& \frac{1}{2}\oint ds\,  {\bf N}^a l_a, \nonumber\\
&=& \frac{1}{2}\oint ds \, {\bar{ \epsilon}}^{ab} l_a \,    {\bf X}    \times {\bf e}_b, \nonumber\\
&=& \frac{1}{2}\oint ds\,    {\bf X}    \times {\bf T}.
\end{eqnarray}

Components of extrinsic curvature are defined as
$
K_{ab}={\bf e}_a\cdot \partial_b {\bf n},
$
where derivatives of the unit normal are 
\begin{eqnarray}
\partial_\theta{\bf n}&=&\left[ \partial_\theta \left(  \frac{r }{\sqrt{r^2 +f'^2}}     \right)   + \frac{ f'}{ \sqrt{r^2+f'^2 }}    \right] {\bf r}\nonumber\\
&+& \left[  \frac{r}{\sqrt{r^2+ f'^2} }  - \partial_\theta \left( \frac{f'}{ \sqrt{r^2+ f'^2} }  \right)   \right]\boldsymbol{\theta},\\
\partial_\phi {\bf n}&=& \frac{r\sin\theta - f'\cos\theta }{ \sqrt{r^2+f'^2 }} \boldsymbol{\phi}.
 \end{eqnarray}
 The calculation of the dot product with the tangent vectors gives the result in the text. We can alternatively made use of
 \begin{equation}
 K_{ab}= ({\bf e}_a \times {\bf e}_c) \cdot (	\partial_b {\bf n} \times {\bf e}^c ).
 \end{equation}
In the same way we can write the gaussian curvature in terms 
of  derivatives of the unit normal as
\begin{eqnarray}
2{\cal R}_G &=& ({\bf e}_a \times {\bf e}_b) \cdot (	\partial^a {\bf n} \times \partial^b {\bf n} ), \nonumber\\
&=&{\cal\varepsilon}_{ab} {\bf n}\cdot (	\partial^a {\bf n} \times \partial^b {\bf n} ),\nonumber\\
&=& 2 \sqrt{g}\, {\bf n}\cdot (	\partial^\theta {\bf n} \times \partial^\phi {\bf n} ).
\end{eqnarray}

\section{Normal integration}
Let us obtain the integral that appears in the Young-Laplace law
\begin{equation}
I= \int dA \, {\bf n}, 
\end{equation}
where $\bf n$ is the unit normal to the vesicle. After substituting its value we have
\begin{eqnarray}
I &=& \int dA \frac{1 }{ \sqrt{r^2 +r'^2 }} ( r\, {\bf r}  - r' \boldsymbol{\theta}), \nonumber\\
&=& 2\pi \int_0^{\theta_0} d\theta \, r   \sin\theta  \, ( r\cos\theta + r'\sin\theta), \nonumber\\
&=& 2\pi \int_0^{\theta_0} d\theta r^2 \sin\theta \cos\theta + 2\pi \int_0^{\theta_0} d\theta r r' \sin^2\theta. \label{I}
\end{eqnarray}
Let us call $I_1$ and $I_2$ the first  and second integrals in \eqref{I} respectively. Thus, up to first order we have 
\begin{eqnarray}
I_1 &=&R^2\int_0^{\theta_0} d\theta [1 + \epsilon f(\theta)]^2\sin\theta\cos\theta, \nonumber\\
&\simeq&R^2 \int_0^{\theta_0}d\theta [1 + 2\epsilon f(\theta)]  \sin\theta\cos\theta,  \nonumber\\
&=&  \frac{ R^2 \sin^2\theta_0}{2}+ 2\epsilon R^2 \int_0^{\theta_0} d\theta f(\theta)  \sin\theta\cos\theta.  \\
I_2&=&  R^2\int_0^{\theta_0} d\theta [1+ \epsilon f(\theta)] \epsilon f'(\theta)\sin^2\theta, \nonumber\\
&\sim& R^2\epsilon \int_0^{\theta_0} d\theta   f'(\theta)\sin^2\theta.
\end{eqnarray}
If we substitute  $x=\cos\theta$ and write the deformation as a Legendre polynomial $f(\theta)=P_l(\theta) $, 
then
\begin{equation}
P'_l =-\sin\theta \dot P_l.
\end{equation}
The integral becomes
\begin{eqnarray}
&& \int_0^{\theta_0} d\theta f(\theta)  \sin\theta\cos\theta = -\int_1^{x_0} \, dx \, x P_l(x) , \nonumber\\
&&=\int_{x_0}^1 dx P_1(x) P_l(x)= \frac{(1-x^2)[ P_l(x)- x\dot P_l(x)] }{2-l(l+1)}, \nonumber\\
&&=\frac{ \sin^2\theta_0  [P_l(\theta)  + \cot\theta_0 P_l' (\theta) ] }{2-l(l+1)},
\end{eqnarray}
where the identity of Legendre functions
\begin{equation}
\int_x^1 dx\,  P_m (x) P_n(x) = \frac{ (1-x^2)[ P_n (x) \dot P_m (x)- P_m(x) \dot P_n(x)  ] }{ m(m+1)- n(n+1) }, 
\end{equation}
has been used.\\
In the same way, we can write
\begin{eqnarray}
&&\int_0^{\theta_0} d\theta   f'(\theta)\sin^2\theta= \int_1^{x_0} dx (1-x^2) \dot P_l (x) .\nonumber\\
&&=  l \int_1^{x_0} dx P_{l-1}(x) - l\int_1^{x_0} dx \,  x P_l (x)\nonumber\\
&&= -l \frac{1-x_0^2 }{l(l-1)}\dot P_{l-1}(x_0) + ... \nonumber\\
&&=  \frac{l \sin\theta }{l(l-1) } P'_{l-1}(\theta)   + l \frac{ \sin^2\theta_0 [ P_l (\theta) + \cot\theta_0 P_l' (\theta)] }{2-l(l+1)},
\end{eqnarray}
in the second line we used the identity
\begin{equation}
(1-x^2) \dot P_l (x)= lP_{l-1}(x) - l\, x P_l(x).  
\end{equation}
Module $2\pi$ and up to first order the integral $I$ turns into
\begin{eqnarray}
I&=&\frac{ R^2 \sin^2\theta_0}{2}+ \epsilon R^2(2+l)  \left[ \frac{ \sin^2\theta_0  [ P_l(\theta)  
+ \cot\theta_0 P_l' (\theta) ] }{2-l(l+1)} \right]\nonumber\\
&&+ \epsilon R^2 \Big[  \frac{ \sin\theta }{(l-1) } P'_{l-1}(\theta) \Big]. 
\end{eqnarray}








\bibliographystyle{abbrv}
\bibliography{refs}

\begin{thebibliography}{99}

\bibitem{diz} A. Diz-Mu\~{n}oz, D. A. Fletcher, O. D. Weiner, {\it Use the force: membrane tension as an organizer of cell shape and motility}, Trends in Cell Biol., {\bf 23}, 47 (2013)
\bibitem{clark} A. G. Clark, O. Wartlick, G. Salbreux, E. K. Paluch, {\it Stresses at the Cell Surface during Animal Cell Morphogenesis}, Curr. Biol. {\bf 24}, R484 (2014).
\bibitem{pontes} B. Pontes, P. Monzo, N. C. Gaithier, {\it Membrane tension: A challenging but universal physical parameter in cell biology}, Sem. In Cell Dev. Biol. {\bf 71}, 30 (2017).
\bibitem{park} Y. K. Park, M. Diez-Silva, G. Popescu, G. Lykotrafitis, W. S. Choi, M. S. Feld and S. Suresh, Proc. Natl. Acad. Sci., {\bf 105}, 13730?13735 (2008).

\bibitem{kunitake} T. Kunitake, {\it Synthetic bilayer membranes: Molecular design, self-organization, and aplplication}, Angew. Chem. Int. Ed. Engl., {\bf 31}, 709-726 (1992).
\bibitem{elani} Y. Elani, R. V. Law, O. Ces, {\it Vesicle-based artificial cells as chemical microreactors with spatially segregated reaction pathways}, Nat. Comms. {\bf 5}, 5305 (2014).
\bibitem{joseph} A. Joseph et al., {\it Chemotactic synthetic vesicles: Design and applications in blood-brain barrier crossing}, Sci. Adv. {\bf 3}, e1700362 (2017).

\bibitem{yuan} H. Yuan, C. Huang, S. Zhang, {\it Dynamic shape transformations of fluid vesicles}, Soft Matter, {\bf 6}, 4571-4579 (2010).
\bibitem{feng} S. Feng, Y. Hu, H. Liang, {\it Entropic elasticity based coarse-grained model of lipid membranes}, J. Chem. Phys.  {\bf 148}, 164705 (2018).

\bibitem{gallop} H.T. McMahon and J.L. Gallop,   {\it Membrane curvature and mechanisms of dynamic cell membrane remodelling}, Nature {\bf 438} 590 (2005).

\bibitem{gueguen} G. Gueguen, N. Destainville, M. Manghi, {\it Fluctuation tension and shape transition of vesicles: renormalisation calculations and Monte Carlo simulations},  Soft Matter, {\bf 13}, 6100, (2017).

\bibitem{fournier} J.B. Fournier, {\it On the stress and torque tensors in fluid membranes}, Soft Matter {\bf 3}, 883 (2007).
\bibitem{galatola} J.B. Fournier and P. Galatola,  {\it Corrections to the Laplace law for vesicle aspiration in 
micropipettes and other confined geometries}, Soft Matter  {\bf 4}, 2463-2470 (2008). 

\bibitem{rozycky} B. R\'{o}\.{z}ycki, R. Lipowsky, {\it Spontaneous curvature of bilayer membranes from molecular simulations:
Asymmetric lipid densities and asymmetric adsorption}, J. Chem. Phys. {\bf 142}, 054101 (2015).

\bibitem{seifert} U. Seifert, {\it Configurations of fluid membranes and vesicles}, Advances in Physics  {\bf 46}:1, 13-137 (1997).

\bibitem{canham} P. B. Canham. {\it The minimum energy of bending as a possible explanation of the biconcave shape of the red
blood cell}. J. Theoret. Biol., {\bf 26}, 61-81, (1970).

\bibitem{helfrich} W. Helfrich, {\it Elastic properties of lipid bilayers-theory and possible experiments}, Z. Naturforsch C {\bf 28}, 
11, 693 (1973).

\bibitem{oyang} Ou-Yang, Zhon-can and W. Helfrich, {\it Instability and deformation of a spherical vesicle by pressure}, Phys. Rev. Lett. {\bf 59}, 2486 (1987).



%
\bibitem{wiese} W. Wiese, W. Helfrich {\it Theory of vesicle budding}, J. Phys.:Condens. Matter {\bf 2} SA329 (1990).
\bibitem{julicher} F. J\"ulicher, R. Lipowsky, {\it Domain induced budding of vesicles}, Phys. Rev. Lett. {\bf 70}, 2964 (1993).
\bibitem{julicher1} F. J\"ulicher, R. Lipowsky, {\it Shape transformations of vesicles with intramembrane domains}, Phys. Rev. E {\bf 53}, 2964 (1996).
\bibitem{miao} L. Miao, U. Seifert, M. Wortis, H.-G. D\"obereiner, {\it Budding transitions of fluid-bilayer vesicles: The effect of area-difference elasticity}. Phys. Rev.  E {\bf 49},  5389 (1994). 
%
\bibitem{agudo} J. Agudo-Canalejo, R. Lipowsky, {\it Stabilization of membrane necks by adhesive particles, substrate surfaces, and constriction forces}, Soft Matter, {\bf 12}, 8155 (2016).
%
\bibitem{dimitrieff} S. Dmitrieff, F. N\'ed\'elec, {\it Membrane Mechanics of Endocytosis in Cells with Turgor}, PLoS Comput. Biol. 11(10), (2015).
%
\bibitem{kas} J. K\"as, E. Sackmann, Biophys. J. {\bf 60}, 825 (1991).
%
\bibitem{golushko1} I. Y. Golushko, S. B. Rochal, {\it Tubular Lipid Membranes Pulled from Vesicles: Dependence of System Equilibrium on Lipid Bilayer Curvature}, Journal of Experimental and Theoretical Physics, {\bf 122}, 169?175, (2016).
\bibitem{shojaei} H. R. Shojaei, M. Muthukumar, {\it Translocation of an Incompressible Vesicle through a Pore}, J. Phys. Chem. B, 120, 6102?6109 (2016).
\bibitem{khunpetch} P. Khunpetch, X. Man, T. Kawakatsu, M. Doi, {\it Translocation of a vesicle through a narrow hole across a membrane}, J. Chem. Phys. {\bf 148}, 134901 (2018).

%
\bibitem{kusu} H. Kusumaatmaja, Y. Li, R. Dimova, R. Lipowsky, {\it Intrinsic Contact Angle of Aqueous Phases at Membranes and Vesicles} PRL 103, 238103 (2009).
\bibitem{li} Y. Li, H. Kusumaatmaja, R. Lipowsky, R. Dimova, {\it Wetting-Induced Budding of Vesicles in Contact with Several Aqueous Phases}, J. Phys. Chem. B, {\bf 116}, 1819?1823, (2012).
%
\bibitem{svetina} B. Bozic, J. Guven, P. Vazquez-Montejo and S. Svetina, {\it Direct and remote constriction of membrane necks}, Phys. Rev. E {\bf 89}, 052701 (2014). 
\bibitem{zia} B. Fourcade, L. Miao, M. Rao, M. Wortis and R.K.P. Zia,  {\it Scaling analysis of narrow necks in curvature models of fluid lipid-bilayer vesicles}, Phys. Rev.  E {\bf 49},  5276 (1994). 
\bibitem{jiang} H. Jiang, G. Huber, R.A. Pelcovits and T. R. Powers, Phys. Rev. E {\bf 76} 031908 (2007).
\bibitem{yang} P. Yang, Q. Du, Z.C. Tu, {\it General neck condition for the limit shape of budding vesicles},
Phys. Rev.  E {\bf 95},  042403 (2017).

\bibitem{jem} R. Capovilla and  J. Guven, {\it Stresses in lipid membranes},  J. Phys. A: Math. Gen. {\bf 35} 
6233 (2002).
\bibitem{barbetta} C. Babetta, A. Imparato and J. B. Fournier, {\it On the surface tension of fluctuating quasi-spherical vesicles}, 
Eur. Phys. J.  E {\bf 31}, 333 (2010). 

\bibitem{willmore} T. J. Willmore, {\it An introduction to  differential geometry} (Dover, New York 1976).

\bibitem{laplace} J. Guven, {\it Laplace pressure as a surface stress in fluid vesicles}, J. Phys. A: Math. Gen. {\bf  4}  

\bibitem{lubarda} V. A. Lubarda,  {\it Mechanics of a liquid drop deposited on a solid substrate}, Soft Matter {\bf 8}, 10288 (2012).

\bibitem{gil} J.A. Santiago, {\it Stresses in curved nematic membranes},  Phys. Rev. E {\bf 97}, 052706 (2018).


 
 \bibitem{giomi-am1} L. Giomi, M. J. Bowick, P. Mishra, R. Sknepnek. and M. C. Marchetti, {\it Defect dynamics in active nematics},  Phil. Trans. R. Soc. A {\bf 372} 20130365 (2014).
 \bibitem{giomi-am2} P. W. Ellis, D. J. G. Pearce, Y.-W. Chang, G. Goldsztein, L. Giomi and A. Fernandez-Nieves, {\it Curvature-induced defect unbinding and dynamics in active nematic toroids
}, Nat. Phys {\bf 14}, 85 (2018).
 \bibitem{alaimo} F. Alaimo, C. K\"ohler, A. Voigt, {\it Curvature controlled defect dynamics in topological active nematics}, Sci. Rep. {\bf 7} 5211 (2017).








\end{thebibliography}

\newpage

\end{document}